\newif\ifformatlipics
\formatlipicsfalse

\ifformatlipics
\documentclass[UKenglish,a4paper]{lipics}
\usepackage{microtype}
\else
\documentclass[11pt]{llncs}
\fi
\usepackage{amsfonts}
\usepackage{amsmath}
\usepackage{calc}
\usepackage{chl-alg}
\usepackage{algorithm}
\usepackage{graphicx}
\newcommand{\dd}{\mathinner{\ldotp\ldotp}}

\newcommand{\BWT}{\ensuremath{\texttt{BWT}}}
\newcommand{\ISA}{\ensuremath{\texttt{ISA}}}
\ifformatlipics
\else
\fi

\ifformatlipics
\title{Low Space External Memory Construction of the Succinct Permuted Longest Common Prefix Array}
\titlerunning{Low Space External Memory Succinct PLCP Array Construction} %optional, in case that the title is too long; the running title should fit into the top page column

\author[1]{German Tischler}
\affil[1]{Max Planck Institute of Molecular Cell Biology and Genetics,\\
	Pfotenhauerstra{\ss}e 108, 01037 Dresden, Germany\\
  \texttt{tischler@mpi-cbg.de}
}
\authorrunning{G. Tischler} %mandatory. First: Use abbreviated first/middle names. Second (only in severe cases): Use first author plus 'et. al.'

\Copyright{German Tischler}%mandatory, please use full first names. LIPIcs license is "CC-BY";  http://creativecommons.org/licenses/by/3.0/

% mandatory: Please choose ACM 1998 classifications from http://www.acm.org/about/class/ccs98-html . E.g., cite as "F.1.1 Models of Computation". 
% mandatory: Please provide 1-5 keywords
\else

\title{Low Space External Memory Construction of the Succinct Permuted Longest Common Prefix Array}
\author{German Tischler}
\institute{%
{Max Planck Institute of Molecular Cell Biology and Genetics,
Pfotenhauerstra{\ss}e 108, 01037 Dresden, Germany}\\
\email {tischler@mpi-cbg.de}
}
\fi

\begin{document}
\maketitle
\begin{abstract}
The longest common prefix (LCP) array is a versatile auxiliary data structure in
indexed string matching. It can be used to speed up searching using the
suffix array (SA) and provides an implicit representation of the topology of an
underlying suffix tree. The LCP array of a string of length $n$ can be represented as an array
of length $n$ words, or, in the presence of the SA, as a bit vector of $2n$ bits
plus asymptotically negligible support data structures. External memory
construction algorithms for the LCP array have been proposed, but those
proposed so far have a space requirement of $O(n)$ words (i.e. $O(n \log n)$
bits) in external memory. This space requirement is in some practical cases prohibitively
expensive. We present an external memory algorithm for constructing the $2n$
bit version of the LCP array which uses $O(n \log \sigma)$ bits of additional space in
external memory when given a (compressed) BWT with alphabet size $\sigma$
and a sampled inverse suffix array at sampling rate $O(\log n)$. This
is often a significant space gain in practice where $\sigma$ is usually much
smaller than $n$ or even constant. We also consider the case of computing
succinct LCP arrays for circular strings.
\end{abstract}
%\vspace{-7mm}
\section{Introduction}
%\vspace{-2mm}
The suffix array (SA) and longest common prefix array (LCP) were introduced as a lower memory variant of the suffix tree
(cf.~\cite{DBLP:conf/focs/Weiner73}) for exact string matching using a pre computed index (cf.~\cite{DBLP:journals/siamcomp/ManberM93}).
For a text of length $n$ both can be computed in linear time in internal memory (IM) (cf.~\cite{DBLP:journals/jacm/KarkkainenSB06,DBLP:conf/cpm/KasaiLAAP01,CHL07cup})
and require $n$ words of memory each.
For large texts the space requirements of SA and LCP in IM can be prohibitive.
Compressed and succinct variants including compressed suffix arrays
(see e.g.~\cite{DBLP:conf/stoc/GrossiV00,DBLP:journals/jal/Sadakane03,DBLP:conf/soda/GrossiGV03}), 
the FM index and variants (see~\cite{DBLP:conf/focs/FerraginaM00,DBLP:journals/isci/FerraginaM01,DBLP:conf/spire/FerraginaMMN04})
and succinct LCP arrays (see \cite{DBLP:journals/mst/Sadakane07}) use less
space, but for practicality it is also crucial to be able to construct these
data structures using affordable space requirements.
Construction algorithms for compressed suffix arrays and the Burrows Wheeler transform (BWT, see \cite{burrows1994block})
using $o(n\log n)$ bits of space in IM (assuming $\sigma\in o(n)$)
were introduced (see e.g.~\cite{DBLP:journals/siamcomp/HonSS09,DBLP:conf/spire/OkanoharaS09}).
It is still unclear whether these algorithms scale well in practice. At the
very least they require an amount of IM which is several times
larger than what is needed for the input text.
External memory solutions for constructing the suffix array and LCP array
have also been presented (see~e.g.\cite{DBLP:conf/alenex/BingmannFO13,DBLP:journals/jea/DementievKMS08,DBLP:conf/wea/KarkkainenK14}).
These algorithms require $O(n)$ words ($O(n\log n)$ bits) of external memory (EM).
However, as for their IM pendants, this space requirement is
large if the algorithms are used as a vehicle to obtain a compressed
representation.
Recently algorithms for constructing the BWT in EM without
explicitly constructing a full suffix array were designed and implemented
(see \cite{DBLP:journals/algorithmica/FerraginaGM12,DBLP:conf/icabd/Tischler14}).
In this paper we present an algorithm for constructing a succinct LCP array
in EM based on a BWT and sampled inverse suffix array
while using $O(n\log\sigma)$ instead of $O(n\log n)$ bits of space in
EM.
Both, BWT and sampled inverse suffix array can be produced in space
$O(n\log\sigma)$ in external memory by the algorithm presented in \cite{DBLP:conf/icabd/Tischler14,mergebwtext}.
In the final section of this paper we consider the extension of our algorithm to circular strings.
%
%\vspace{-2mm}
\section{Definitions}
%\vspace{-2mm}
%
Let $\Sigma$ denote a totally ordered and ranked alphabet, w.l.o.g. we assume
$\Sigma=\{0,1,\ldots,\sigma-1\}$ for some $\sigma>0$. Further let $s=s_0s_1\ldots s_{n-1}$ denote
a string of length $|s|=n>0$ over $\Sigma$ s.t.~the last symbol of $s$ is the
minimal symbol in $s$ and does not appear elsewhere in $s$. We use $s[i]$ to denote $s_i$
and $s[i\dd j]$ for $s_is_{i+1}\ldots s_j$ for $0\leq i \leq j < n$. 
$s[i\dd j]$ denotes the empty string for $i>j$.
The $i$'th suffix of $s$ denoted by $\tilde{s}_i$ is the string $s[i\dd n-1]$. Suffix $\tilde{s}_i$ is smaller than
$\tilde{s}_j$ for $i\ne j$ (denoted by $\tilde{s}_i < \tilde{s}_j$) if for the smallest
$k$ s.t.~$s[i+k]\ne s[j+k]$ we have $s[i+k]<s[j+k]$. The suffix array \texttt{SA} of $s$
is the permutation of $0,1,\ldots,n-1$ s.t.~$\tilde{s}_{\texttt{SA}[i-1]} < \tilde{s}_{\texttt{SA}[i]}$
for $i=1,2,\ldots,n-1$. For two suffixes $\tilde{s}_i$ and $\tilde{s}_j$
with $i\ne j$ the longest common prefix $\texttt{lcp}(i,j)$ of the two is $s[i\dd i+\ell-1]$
for the smallest $\ell$ s.t.~$s[i+\ell]\ne s[j+\ell]$. The array \texttt{LCP}
of $s$ is defined by $\texttt{LCP}[i] = |\texttt{lcp}(\texttt{SA}[i-1],\texttt{SA}[i])|$
for $i>0$ and $\texttt{LCP}[0]=0$. The inverse suffix array $\texttt{ISA}$
of $s$ is defined by $\texttt{ISA}[\texttt{SA}[i]]=i$ for $0\leq i<n$. The
permuted LCP array $\texttt{PLCP}$ of $s$ is given by $\texttt{PLCP}[i]=\texttt{LCP}[\texttt{ISA}[i]]$
for $0\leq i<n$ and $\texttt{PLCP}[i]=0$ otherwise. 
The Burrows Wheeler transform $\texttt{BWT}$ of $s$ is defined by $\texttt{BWT}[i]=s[(\texttt{SA}[i]+n-1)\textnormal{ mod }n]$ for $0\leq i<n$.
Let $C$ be the array of length $\sigma$ s.t.~$C[a]=|\{i\mid  s[i] = a\}|$
for $a\in\Sigma$ and let $D$ be an array of length $\sigma+1$ s.t.~$D[a]=\sum_{i<a} C[i]$
for $0\leq a\leq\sigma$. For a sequence $t=t_0,t_1,\ldots, t_{k-1}$ for some $k\geq 0$ let
$\textsc{rank}_t(a,j)=|\{i|0 \leq i < \min(j,k), t_i=a\}|$, i.e.~the number of $a$
elements in $t$ up to but excluding index $j$ and let
$\textsc{select}_t(a,j)=\min\{i\mid \textsc{rank}_t(a,i+1)=j+1\}$
if $0\leq j < \textsc{rank}_t(a,k)$ and undefined otherwise. $\textsc{LF}$ is
defined by $\textsc{LF}(r)=\texttt{ISA}[(\texttt{SA}[r]+n-1)\textnormal{ mod }n]$.
$\textsc{B}$ is defined by $\textsc{B}(a,i)=D[a] + \textsc{rank}_{\texttt{BWT}}(a,i)$
for $a\in\sigma, 0\leq i\leq n$ and $\textsc{backstep}$ by
$\textsc{backstep}(a,(i,j))=(\textsc{B}(a,i),\textsc{B}(a,j))$ for
$a\in \Sigma, 0\leq i,j \leq n$.
%
%\vspace{-2mm}
\section{Previous Work}
%\vspace{-2mm}
%
The first linear time algorithm for computing the LCP array from the suffix
array and text appeared in \cite{DBLP:conf/cpm/KasaiLAAP01}. One of the main
combinatorial properties used by this algorithm is the fact that $\texttt{PLCP}[i]\geq \texttt{PLCP}[i-1]-1$ for
$0<i<n$. This property is also used in \cite{DBLP:journals/mst/Sadakane07} to
obtain a representation of the PLCP array using $2n+o(n)$ bits while
allowing constant time access. Let $\zeta(0)=1$ and $\zeta(i)=0\zeta(i-1)$
for $i > 0$. The $2n$ bits in the data structure are the bit sequence
$K = \eta(n-1)$ given by $\eta(0)=\zeta(\texttt{PLCP}[0]+1)$ and $\eta(i)=\eta(i-1)\zeta(\texttt{PLCP}[i]-\texttt{PLCP}[i-1]+1)$ for $0<i<n$.
The $o(n)$ additional bits are used for a select index (cf.\cite{DBLP:conf/fsttcs/Munro96})
on $K$. $K$ stores the sequence of pairwise differences of adjacent
PLCP values shifted by $1$ in unary representation (the number $i$ is
represented as $i$ zero bits followed by a $1$ bit). The value
$\texttt{PLCP}[i]$ can be retrieved as $\textsc{select}_K(1,i)-2(i+1)-1$.
In \cite{DBLP:journals/jda/BellerGOS13} Beller et al present an algorithm
for computing the LCP array in IM using a wavelet tree (see
\cite{DBLP:conf/soda/GrossiGV03}). This algorithm runs for $\ell_m+1$ rounds
where $\ell_m$ is the maximum LCP value produced. Round $i$ for $0\leq i\leq \ell_m$ sets
$\texttt{LCP}[r]$ for exactly those ranks $r$ s.t.~$\texttt{LCP}[r]=i$,
i.e.~the values are produced in increasing order. 
%
%\vspace{-2mm}
\section{Computing the succinct \texttt{PLCP} array}
%\vspace{-2mm}
\label{succinctlcpalgorithm}
In this section we modify the algorithm by Beller et al (cf.~\cite{DBLP:journals/jda/BellerGOS13})
to produce the succinct 2n bit PLCP bit vector in EM. The main idea
is to use the fact that the algorithm produces the LCP values in increasing
order. It starts with a tuple $(\epsilon,(0,n))$ which denotes the empty
word and the corresponding rank interval on the suffix array (the lower end
$0$ is included, the upper $n$ is excluded). Round $i$
takes the tuples from the previous round (or the start tuple for round $0$)
and considers all possible extensions by one symbol via backward search
(cf.~\cite{DBLP:conf/focs/FerraginaM00}), i.e.~it produces $(aw,(l',r'))$ from $(w,(l,r))$ for each $aw$ appearing in $s$.
All suffixes considered in round $i$ starting by $aw$ in the rank interval $(l',r')$ 
have a common prefix of length $i+1$, while the suffixes at ranks $l'-1$ and
$l'$ (for $l\ne 0$) as well as at ranks $r'-1$ and $r'$ (for $r'<n$) have a
common prefix of at most length $i$. Based on this insight we can set
$\texttt{LCP}[l']$ and $\texttt{LCP}[r']$ to $i$, if they have not already
been set in a previous round. In the tuples the first (string) component is only provided
for the sake of exposition, the algorithm does not require or use it. In
addition the algorithm prunes away intervals when a respective LCP value
(Beller et al use the upper bound $r'$ for setting new values in
\cite{DBLP:journals/jda/BellerGOS13}, we in this paper use the lower bound $l'$ as it simplifies the transition to EM)
is already set.

The succinct PLCP array $K$ contains $n$ zero and $n$ one bits. The one bits
mark positions in the text (remember PLCP is in text order). 
The zero bits encode the differences between
adjacent PLCP values shifted by $1$. For computing this bit vector assume
that we start off with a vector of $n$ one bits. The information we need in
addition is in front of which $1$ bit we have to insert how many $0$ bits.
If $\texttt{PLCP}[i]$ is not smaller than $\texttt{PLCP}[i-1]$, then we have
to add $\texttt{PLCP}[i] - \texttt{PLCP}[i-1]+1$ zero bits just in front of
the $i+1$st $1$ bit.
In the algorithm we can achieve this by starting to add
$0$ bits for ranks which did not have their value set in a previous round but which
do have the value for the rank of the previous position set in the current
round. We call this adding a rank to the active set.
We stop adding $0$ bits for a rank in the round in which the value for the rank itself gets
set, which we call removing a rank from the active set. Figure \ref{intmemalg}
shows an algorithm implementing this approach in IM. 
%$\texttt{W}$ is a wavelet tree representing the $\texttt{BWT}$ of $s$.
A wavelet tree (cf.~\cite{DBLP:conf/soda/GrossiGV03}) for \texttt{BWT} 
can be used to compute the $\textsc{backstep}$, $\textsc{rank}$ and
$\textsc{select}$ functions in time $O(\log\sigma)$ and to determine the
set of symbols occurring in any index interval on $\texttt{BWT}$ in time $O(\log\sigma+o)$
where $o$ is the number of distinct symbols in the interval.
\begin{figure}
\begin{algo}{PLCPinternal}{\texttt{BWT},\texttt{n},\texttt{ISA}}
	\SET{(\texttt{Q},\texttt{activeSet},\texttt{S},\texttt{PD})}{(\emptyset,\emptyset,\emptyset,\emptyset)}
	\ACT{\texttt{Q}.\textsc{enque}((0,\texttt{n}))}
	\DOWHILE{\texttt{Q}.\textsc{empty}() = \textsc{false}}
		\COM{Queue for next round and ranks set in this round}
		\SET{(\texttt{NQ},\texttt{T})}{(\emptyset,\emptyset)}
		\DOWHILE{\texttt{Q}.\textsc{hasNext}()}
			\SET{(\texttt{l},\texttt{r})}{\texttt{Q}.\textsc{next}()}
			\DOFOREACH{\texttt{sym} \in \{ \texttt{BWT}[i] \mid \texttt{l} \leq i < \texttt{r} \}}
				\SET{(\texttt{l'},\texttt{r'})}{\textsc{backstep}(\texttt{sym},(\texttt{l},\texttt{r}))}
				\IF{\texttt{S}.\textsc{contains}(\texttt{l'}) = \textsc{false}}
					\COM{get \texttt{l'src} s.t.~\textsc{LF}(\texttt{l'src})=\texttt{l'}}
					\COM{this is the smallest $i$ s.t.~$l\leq i<r$ and \texttt{BWT}[i]=\texttt{sym}}
					\SET{\texttt{l'src}}{\textsc{select}_{\texttt{BWT}}(\texttt{sym},\textsc{rank}_{\texttt{BWT}}(\texttt{sym},\texttt{l}))}
					\COM{mark \texttt{l'} as to be set in this round}
					\ACT{\texttt{T}.\textsc{insert}(\texttt{l'})}
					\COM{put \texttt{l'src} in active set if not set yet}
					\IF{\texttt{S}.\textsc{contains}(\texttt{l'src}) = \textsc{false}}
						\ACT{\texttt{activeSet}.\textsc{insert}(\texttt{l'src})}
					\FI
					\ACT{\texttt{NQ}.\textsc{enque}((\texttt{l'},\texttt{r'}))}
				\FI
			\OD
		\OD
		\COM{Increment number of $0$ bits for ranks in active set}
		\DOFOREACH{\texttt{r}\in \texttt{activeSet}}
			\IF{\texttt{PD}.\textsc{contains}(\texttt{r})}
				\SET{\texttt{PD}[\texttt{r}]}{\texttt{PD}[\texttt{r}]+1}
			\ELSE
				\SET{\textnormal{ }\texttt{PD}[\texttt{r}]}{1}
			\FI
		\OD
		\COM{Remove ranks set in this round from active list}
		\COM{and update set of ranks finished}
		\DOFOREACH{\texttt{r}\in \texttt{T}}
			\IF{\texttt{activeSet}.\textsc{contains}(\texttt{r})}
				\ACT{\texttt{activeSet}.\textsc{remove}(\texttt{r})}
			\FI
			\ACT{\texttt{S}.\textsc{insert}(r)}
		\OD
		\SET{\texttt{Q}}{\texttt{NQ}}
	\OD
	\COM{Produce succinct bit vector in text order}
	\SET{(i,\texttt{K})}{(0,\emptyset)}
	\DOFORI{p}{0}{n-1}
		\SET{r}{\texttt{ISA}[\texttt{p}]}
		\IF{\texttt{PD}.\textsc{contains}(\texttt{r})}
			\DOFORI{j}{1}{\texttt{PD}[\texttt{r}]}
				\SET{\texttt{K}[\texttt{i++}]}{\textsc{false}}
			\OD
		\FI
		\SET{\texttt{K}[\texttt{i++}]}{\textsc{true}}
	\OD
	\RETURN{B}
\end{algo}
\caption{Internal memory version of PLCP computation algorithm}
\label{intmemalg}
\end{figure}

In the following we show how to adapt this algorithm so it becomes usable in
EM and requires no more than $O(n\log\sigma)$ space in EM
while using $O(\sigma\log n)$ bits of IM. This means we
need to make sure that all data structures used in EM are accessed in a 
purely sequential way and none use $\omega(n\log\sigma)$ space. In particular
we need to consider the representation and access patterns of the queues \texttt{Q} and
\texttt{NQ}, the Burrows Wheeler transform \texttt{BWT} of $s$, the sets \texttt{S}, \texttt{T}
and \texttt{activeSet} and the counter array for zero bits \texttt{PD}.

For some of the representations we will use Elias $\gamma$ code (cf. \cite{DBLP:journals/tit/Elias75})
and the following result proven in \cite{mergebwtext}.
\begin{lemma}[\cite{mergebwtext}]
\label{gammagapfulllemma}
Let $G$ denote an array of length $\ell$ such that $G[i]\in\mathbb{N}$ for
$0\leq i < \ell$ and $\sum_{i=0}^{l-1} G[i]=s$ for some $s\in\mathbb{N}$. Then the $\gamma$
code for $G$ takes $O(\ell + s)$ bits.
\end{lemma}
This means we can represent any strictly increasing sequence
$x_0,x_1,\ldots,\allowbreak x_{k-1}$ of numbers from ${0,1,\ldots,N}$ for $N\in O(n)$ and $k>0,k\in O(n)$ in $O(n)$
bits by storing the differences $x_i-x_{i-1}$ for $i=0,1,\ldots,k-1$ in $\gamma$ code were we
assume $x_{-1}=-1$.

\begin{itemize}
\item The queue \texttt{NQ} is not produced in increasing order in the algorithm
as stated in Figure \ref{intmemalg} (meaning if $(l_1,r_1)$ is enqueued right
after $(l_0,r_0)$ then we cannot assume $l_1\geq r_0$). If however the queue $Q$ is in
increasing order and we consider only the intervals produced by extensions
with a fixed symbol \texttt{sym}, then those extension intervals are in increasing order.
The $\textsc{B}$ function is for a fixed first argument $\texttt{sym}$ monotonously increasing in it's second argument
and has a maximum value of $\texttt{D}[\texttt{sym}+1]$ which is only
reached as an (excluded) right end of any \textsc{backstep} call and at the
same time the (included) minimum left end of calls for \textsc{backstep}
with first parameter \texttt{sym+1} (if any such exist in $s$).
This means if we replace $\texttt{NQ}.\texttt{enque}((\texttt{l}',\texttt{r}'))$ by
$\texttt{NQ}.\textsc{enque}(\texttt{sym},(\texttt{l}',\texttt{r}'))$, sort
$\texttt{NQ}$ stably by the first (\texttt{sym}) component and subsequently drop the
first component then the resulting list of intervals will be in sorted
order. The sorting can be performed using $O(\log{\sigma})$ rounds of bucket
sorting along the bit representation of the first component, each of which takes $O(n)$ time as we can never have more than $n$
elements in the queue. During the whole sorting procedure the elements for
each single first component will stay in ascending order concerning their
second component, which allows us to store the second component using differential
$\gamma$ code. The sequence of lower interval bounds and the one of upper
interval bounds both form strictly increasing sequences. Starting the
difference coding for the sequences for \texttt{sym} at
$\texttt{D}[\texttt{sym}]-1$ ensures that for both sequences the sum of the
stored numbers does not exceed $n$, so we can store them using $O(n)$ bits
according to Lemma \ref{gammagapfulllemma}.
\item The $\texttt{T}$ set stores a subset of the lower interval bounds produced
for $\texttt{NQ}$. We can thus use similar steps to produce it in sorted
order while requiring $O(n\log\sigma)$ bits of space in EM and
$O(\sigma\log n)$ bit of IM.
\item The values added to \texttt{activeSet} in line $18$ can easily be added
in increasing order by first storing them in a heap data structure
for each source interval $(\texttt{l},\texttt{r})$ and writing the values out in order at the
end of the handling of $(\texttt{l},\texttt{r})$.
This takes space $O(\sigma\log n)$ in IM while the run time for this is bounded
by $O(n\log \sigma)$ for each round (the heap depth is bounded by $\log\sigma$
as we never insert more than $\sigma$ elements into any heap and the total number of elements added is bounded by $n$).
The values in increasing order can again be stored using differential
$\gamma$ code in $O(n)$ bits.
As soon as we have the set of newly added values for a round we can merge it into the set of previously added values,
which can be stored in the same way.
Storing \texttt{activeSet} in this way requires $O(n)$ bits of space in EM.
\item
For each source interval $(\texttt{l},\texttt{r})$
the set of symbols in $\{\texttt{BWT}[i]\mid \texttt{l}\leq i < \texttt{r}\}$, 
the target intervals $(\texttt{l}',\texttt{r}')$ and the respective $\texttt{l'src}$
values can be computed during a linear scan of the $\texttt{BWT}$ sequence
streamed from EM while keeping track of the values of the
$\textsc{rank}$ function for each symbol.
This requires $O(\sigma\log n)$ bits of space in IM. 
We keep tuples $(\texttt{sym},\texttt{l'},\hat{\texttt{r}},\texttt{l'src})$ in an AVL tree
(cf.~\cite{tree-adelson-62}) where only the first ($\texttt{sym}$) component
is used as the key. While scanning $\texttt{BWT}$ we insert
$(\texttt{sym},\textsc{B}(\texttt{sym},\texttt{l'src}),\textsc{B}(\texttt{sym},\texttt{l'src})+1,\texttt{l'src})$ upon first
encountering $\texttt{sym}$ at index $\texttt{l'src}$ in $(\texttt{l},\texttt{r})$
and update the third component accordingly whenever we find another instance of $\texttt{sym}$ in the source interval. 
With the same reasoning as above for the heap used while handling \texttt{activeSet} this takes time $O(n\log \sigma)$ for one round.
\item The accesses to $\texttt{S}$ in line $17$ are in ascending index order
and updating $\texttt{S}$ in line $30$ while scanning $\texttt{S}$ and
$\texttt{T}$ can read both sequences in linear ascending order, which is
suitable for EM.
Accessing $\texttt{S}$ at $\texttt{l'}$ in line $10$ is somewhat more challenging. As shown above the $\texttt{l}'$
values in each round are only increasing when we look at a single symbol
$\texttt{sym}$.
We can obtain the bits we need to see in the required order
using the following steps.
First compute the sequence of $\texttt{l}'$ values we
need to access in ascending order. This can be done as described above for
producing $\texttt{NQ}$, i.e.~produce a set of pairs
$(\texttt{sym},\texttt{l}')$, sort it by the first component while using differential $\gamma$ code
for representing the second components and then drop the first component.
This takes time $O(n\log\sigma)$ and space $O(n\log\sigma)$ bits in EM. It gives us the set of required
$\texttt{l}'$ values in increasing order and thus makes it easy to determine
whether $\texttt{S}$ does or does not contain the respective values, which
we store as a bit vector in EM.
This bit vector has as many bits as $\texttt{l}'$ values relevant in the current round, which is $O(n)$.
Now we have the relevant bits, but they are in the wrong order, as we sorted
the $\texttt{l}'$ values by the respective $\texttt{sym}$ values.
We can reorder the bits by {\em inverse sorting} them using the original order of
the $\texttt{sym}$ values.
Figure \ref{binUnBucketSort} shows an algorithm which performs inverse sorting of a sequence $\texttt{A}$ given a binary key
vector $\mathcal{K}$.
\begin{figure}[t]
\begin{algo}{binUnBucketSort}{\mathcal{K},\texttt{A},\texttt{m}}
	\SET{(\texttt{cnt}[0],\texttt{cnt}[1])}{(0,0)}
	\DOFORI{i}{0}{\texttt{m}-1}
		\SET{\texttt{cnt}[\mathcal{K}[i]]}{\texttt{cnt}[\mathcal{K}[i]]+1}
	\OD
	\SET{(\texttt{cnt}[0],\texttt{cnt}[1])}{(0,\texttt{cnt}[1])}
	\DOFORI{i}{0}{\texttt{m}-1}
		\SET{\texttt{B}[i]}{\texttt{A}[\texttt{cnt}[\mathcal{K}[i]]]}
		\SET{\texttt{cnt}[\mathcal{K}[i]]}{\texttt{cnt}[\mathcal{K}[i]]+1}
	\OD
	\RETURN{B}
\end{algo}
\caption{Inverse binary bucket sorting for key vector $\mathcal{K}$ and data vector A, both of length $m$}
\label{binUnBucketSort}
\end{figure}
It does this by first determining how many $0$ and $1$ bits there are in the key vector (lines 1-3) and then rebuilding the original
sequence by scanning $\mathcal{K}$ and taking elements from the $0$ and $1$
regions of the sorted sequence in accordance with the key bits encountered
(lines 5-7).
This inverse binary bucket sorting can be extended to inverse radix sorting for non binary keys.
It requires time $O(n\log\sigma)$ (we need $\log\sigma$ rounds of inverse bucket sorting) and space $O(n\log\sigma)$ bits in EM.
\item The $\texttt{PD}$ array can be represented as a bit vector in EM. 
We initialise it as a vector of $n$ one bits. Adding one to index
$r$ is done by inserting a zero bit just ahead of the $k+1$'st one bit.
We scan $\texttt{activeSet}$ and $\texttt{PD}$ linearly for updating $\texttt{PD}$
where $\texttt{PD}$ has at most $2n$ bits at any time. So updating
$\texttt{PD}$ in each round takes $O(n)$ time and storing $\texttt{PD}$
takes $O(n)$ bits in EM.
\end{itemize}
Overall each round of the algorithm up to line $32$ takes time $O(n\log\sigma)$ and we need
$O(n\log\sigma)$ bits of space in EM. In the worst case the
maximum LCP value is $n-2$ (which is e.g.~reached $s[i]=1$ for $0\leq i<n-1$ and $s[n-1]=0$), 
so the worst case run time of the algorithm is $O(n^2\log\sigma)$. In the average case (cf.~\cite{szpankowski1991height})
the maximum value is in $O(\log_\sigma n)$, which gives this part
of the algorithm a run time of $O(n\log n\allowbreak \log\sigma)$ on average.
This leaves us with the issue that the procedure above so far produces the
difference between PLCP values in rank instead of position order. This is
set right by lines 33-39 in Figure \ref{intmemalg}, however it uses a
complete inverse suffix array and requires random access to the PD array.
Given a sampled inverse suffix array at sampling rate $\mathcal{s}\in O(\log n)$ taking
$O(n)$ bits, the \texttt{BWT} and the $\texttt{PD}$ bit vector we can
produce the final PLCP bit vector using the following steps:
\begin{enumerate}
\item Create pairs $(\texttt{ISA}[i\mathcal{s}],\texttt{i}\mathcal{s})$
for $i=0,1,\ldots,\lceil \frac{n}{ \mathcal{s} } \rceil-1$ in EM
 from the sampled inverse suffix array (both components are stored as $O(\log n)$ bit block code)
and sort these pairs by their first
(rank) component using radix sort. This takes space $O(n)$ bits in EM and time $O(\frac{n}{\mathcal{s}}\log n)=O(n)$.
After sorting annotate each tuple with one bit set to \texttt{true} as third
component (marks the tuple as active), the number $0$ stored in $\gamma$
code as fourth component (stores the number of PLCP values added to the tuple so far) and an empty vector of $\gamma$ coded numbers as the fifth component.
\item For $\mathcal{s}$ rounds do the following: perform an \textsc{LF}
operation on the tuples (map $(r,p,a,b,c)$ to $(\texttt{BWT}[\texttt{r}],\textsc{LF}(r),p',a,b,c)$ where $p'=(p+n-1)\textnormal{ mod }n$ if $a$ is \texttt{true} and $p$ otherwise)
by scanning the $\texttt{BWT}$ and computing $\textsc{LF}$ as described
above while tracking the $\textsc{B}$ function using $O(\sigma\log n)$ bits of IM.
Sort the resulting tuples by the first component and drop the first
component. This restores the sorted order according to the rank of the tuples
and takes time $O(n\log\sigma)$ and space $O(n\log\sigma)$ bits in EM.
Note that for each active (third component is \texttt{true}) 
tuple in the list we retain the invariant that for a first component $r$ we
have $\texttt{SA}[r]$ as the second component.
Scan the tuples and the $\texttt{PD}$ bit vector and copy the respective
(matching rank) values into tuples marked as active by inserting the value $\texttt{PD}(r)$
at the front of the vector of $\gamma$ coded values in component five and incrementing the counter
for appended values (fourth component) by one. This takes time $O(n)$ and
again space $O(n\log\sigma)$ bits in EM. In another scan mark
tuples s.t.~their second component $p$ is divided by $\mathcal{s}$ as inactive.
Note that at the end of each round we have the following property:
Let $(r,p,a,c,(v_0,v_1,\ldots,v_{c-1}))$ be a tuple in our list. Then for
$i=0,1,\ldots,c-1$ we have $v_i = \texttt{PLCP}[p+i]-\texttt{PLCP}[p+i-1]+1$.
\item Sort the tuples by the second component (position) using a $\log n$ round radix sort taking $O(n)$ time and $O(n\log\sigma)$ bits of space.
Let $t_0,t_1,\ldots,t_{\lceil\frac{n}{\mathcal{s}}\rceil-1}$ be the sequence
of tuples we have obtained. Then for each $t_i=(r,p,a,b,c)$ with $0\leq i
\leq \lceil\frac{n}{\mathcal{s}}\rceil-1$
we now have $p=i\mathcal{s}$, $a=\texttt{false}$, $b$ represents $\min(n-p,\mathcal{s})$
and $c$ is the sequence $v_0,v_1,\ldots,v_{b-1}$ s.t.~$v_i = \texttt{PLCP}[p+i]-\texttt{PLCP}[p+i-1]+1$.
\item Initialise an empty bit vector $K$. Scan the tuples and for each
tuple do the following: let $c$ denote the number stored in the fourth
component and let $v_0,v_1,\ldots,v_{c-1}$ be the (decoded) numbers stored
in the fifth component. For $i$ in $0,1,\ldots,c-1$ append $v_{i}$ 
zero bits to $K$ and then 1 one bit.
\end{enumerate}
The bit vector $K$ is by construction the succinct $2n$ bit representation
of the PLCP array. The whole reordering takes time $O(n\log n\log\sigma)$,
$O(n\log\sigma)$ bits of space in EM and $O(\sigma\log n)$ bits
of space in IM. 
Each tuple at maximum uses $\log\sigma$ bits for the symbol intermediately
introduced in step $2$, $O(\log n)$ bits for rank and position and $O(\log n)$
bits for storing the number of $\texttt{PD}$ values copied into the tuple so far.
The sum over all stored $\gamma$ values in the last component of the tuples
is bounded by $n$ and reaches $n$ at the end of the procedure. 
We summarise the run time and space requirements of the EM algorithm in the following Theorem.
\begin{theorem}
The succinct 2n bit PLCP representation for a string $s$ of length $n$ can,
given it's \texttt{BWT} and sampled suffix array of sampling rate $\mathcal{s}\in O(\log n)$, be constructed in worst cast time $O(n^2\log\sigma)$
and average time $O(n\log n\allowbreak\log\sigma)$ using $O(n\log \sigma)$ bits of space in EM and $O(\sigma\log n)$ bits of space in IM.
\end{theorem}
\section{Reducing Internal Memory Usage}
\label{redimsect}
While the algorithm of the previous section has space requirements in
$O(n\log\sigma)$ bits in external memory, the need for $O(\sigma\log n)$
bits in IM may be considered as too large in some situations, even though it
is not an obstacle in practice. We can modify the algorithm to use less
space in internal memory, as we show in the following. A suitable
reformulation of the algorithm is given in Figure \ref{lowimalg}. The
algorithm as shown only reformulates the computation of the bit vector up to
the point were it is translated from rank to position order. The crucial
point about the reformulation is to compute the \textsc{LF} and
\textsc{backstep} functions without keeping track of the value of the
\textsc{rank} function in IM for each single symbol in $\Sigma$. Observe
that given a set of ranks $R$ we can compute the set of ranks $R_{\textsc{LF}}$
defined by $R_{\textsc{LF}}=\{r' \mid r' = \textsc{LF}(r),~r\in R\}$ using
the following steps: create a bit vector $R_B$ of length $n$ s.t.~${R_B}_r=1$
iff $r\in R$ and then construct the sequence of pairs $P_R=(\BWT_0,{R_B}_0)(\BWT_1,{R_B}_1)\ldots (\BWT_{n-1},{R_B}_{n-1})$.
Sort $P_R$ by the first (symbol) component in a stable way using radix
sort in time $O(n\log\sigma)$. It is easy to see that the second (bit) component
of the sorted vector represents $R_{\textsc{LF}}$ by virtue of marking the
respective ranks by $1$ bits. This method can be extended to computing the
\textsc{backstep} function for a given set of intervals and all possible
extensions of the respective intervals on the left. To this end observe that
for a given interval $[l,r)$ an extension is possible by exactly those
symbols contained in the set given by $\{a \mid a=\BWT_i \textnormal{ for some }l=\leq i < r\}$, 
the lower bound $l'$ of $(l',r')=\textsc{backstep}(a,(l,r))$ for any such
symbol is given by $l' = \textsc{LF}(l_{src})$ where $l'$ is the smallest
number s.t.~$l\leq l_{src} < r$ and $\BWT_{l_{src}}=a$ and $r'-l'$ equals
the number of $a$ symbols in the sequence $\BWT_l,\BWT_{l+1},\ldots,\BWT_{r-1}$.
The depicted algorithm computes all extensions of a given set of intervals by the $\textsc{backstep}$
function using the following steps.
Assume a list of intervals $L=(l_0,r_0), (l_1,r_1)\ldots,(l_{m-1},r_{m-1})$
is given s.t.~$l_0=0$, $r_{i-1}=l_{i}$ for $i=1,2,\ldots,m-1$ and $r_{m-1}=n$.
In particular the intervals partition the index space $0,1,\ldots,n-1$.
$L$ can be stored using $O(n)$ bits in external memory using either $\gamma$
code for storing the increasing sequences of lower and upper bounds using
differential encoding or by storing two bit vectors of length $n$ marking
the start and end of the intervals.
For each interval $(l_i,r_i)$ in ascending order do the following to produce a sequence $\mathcal{Z}$:
\begin{enumerate}
\item extract the sequence $B=\BWT_{l},\BWT_{l+1},\ldots,\BWT_{r-1}$ to $B_S$ and sort it in time $O((r-l)\log \sigma)$ using radix sort
\item in a single linear scan of $B_S$ mark the first occurance of symbol
$a$ in $B_S$ with the number of times it occurs in $B_S$, i.e. 
$|\{i \mid 0 \leq i < r-l\textnormal{ and } {B_S}_i = a\}| = |\{i \mid 0\leq i < r-l \textnormal{ and } B_i = a\}| = |\{i \mid l \leq i < r \textnormal{ and }\BWT_i =
a\}|$. The rest of the character instances are marked with zero. The attached numbers are stored using $\gamma$ code. 
The numbers stored obviously sum up to $r-l$. Let the obtained sequence be called $B_M$.
\item append $B_M$ to $\mathcal{Z}$.
\end{enumerate}
Then sort $\mathcal{Z}$ stably by the first (symbol) component using radix sort in time $O(n\log\sigma)$.
Let $\mathcal{Z}_S = (a_0,v_0),(a_1,v_1),\ldots,(a_n,v_n)$ denote the resulting sorted sequence.
Further let $J=\{ j \mid v_j \ne 0\} = j_0, j_1, \ldots, j_{k-1}$ and
$I=(j_0,v_{j_0}),(j_1,v_{j_1}),\ldots,(j_{k-1},v_{k-1})$. 
Let \[\textsc{backstep}^\ast(a,L)=\textsc{backstep}(a,(l_0,r_0)),\ldots,\textsc{backstep}(a,(l_{m-1},r_{m-1}))\]
for $a\in\Sigma$ and
\[\textsc{backstep}'(L)=\textsc{backstep}^\ast(0,L),\ldots,\textsc{backstep}^\ast(\sigma-1,L)\enspace .\]
Let the filter function $\textsc{flt}$ be defined by
\[
\textsc{flt}((\alpha_1,\beta_1),(\alpha_2,\beta_2),\ldots,(\alpha_{z},\beta_{z}))=
	\left\lbrace
	\begin{array}{ll}
	(\alpha_1,\beta_1),\textsc{flt}((\alpha_2,\beta_2),\ldots,(\alpha_{z},\beta_{z})) & \textnormal{if }\alpha_1\ne\beta_1\\
	\textsc{flt}((\alpha_2,\beta_2),\ldots,(\alpha_{z},\beta_{z})) & \textnormal{otherwise}\\
	\end{array}
	\right.
\]
Following the same pattern as computing the $\textsc{LF}$ function by attaching the BWT symbols
to a bit vector it is straight forward to see that $I$ is exactly the
sequence of intervals $\textsc{flt}(\textsc{backstep}'(L))$, i.e.~all
non empty extensions of intervals in $L$ in ascending order. In consequence
we obtain the following result.
\begin{lemma}
Given $\BWT$ and a sorted, non overlapping list of intervals $L$ drawn from $[0,n)$
s.t.~both $\BWT$ and $L$ can be decoded in constant time per element
the sorted sequence of intervals $\textsc{flt}(\textsc{backstep}'(L))$ can
be computed in time $O(n\log\sigma)$ and space $O(n\log\sigma)$ bits in EM
and $O(\log n + \log\sigma)$ in IM.
\end{lemma}
In each round we activate ranks $r$ s.t.~$\textsc{LF}(r)$ gets set in this
round while $r$ itself has not already been set in a previous round.
We keep a bit vector $S$ in external memory marking the indices of ranks for
which we already observed the corresponding LCP value in a previous round.
Remember that a rank $l'$ gets set on $S$ in the first round in which
$l'$ appears as a result interval lower bound of a call to $\textsc{backstep}(a,(l,r))$ for any arguments $a,l$ and r.
The result intervals for the $\textsc{backstep}$ operation are encoded in
the sequence $\mathcal{Z}$ in the algorithm in Figure \ref{lowimalg} after
it has been sorted in line $21$. Interval start points are marked by such
tuples which have a non zero count (second component) attached. The
information whether or not a rank will be newly set in $S$ in the current
round is encoded in the sequence $\mathcal{Z}'$ in lines $22-25$ of the
algorithm. We perform an inverse \textsc{LF} mapping on $\mathcal{Z}'$ by
performing an inverse sorting of $\mathcal{Z}'$ using $\textsc{BWT}$ as key
sequence. This allows us to determine which ranks need to be activated by
combining information from the sequence $S$ and $\mathcal{Z}'$ during a
linear scan of the two sequences (lines $28-32$). The active set can be
stored as a bit vector marking active ranks. The algorithm produces the
indices of newly activated ranks in increasing order, so merging them into
the already existing set is trivially performed in linear time $O(n)$. 
We keep the encoding of the $\texttt{PD}$ vector from the previous section. 
Updating it by incrementing the counts for active ranks is straight forward
and takes time $O(n)$. Finally the algorithm cleans the active set, sets
the new ranks in $S$ and computes the input intervals for the next round in
lines $34-41$. Again all of this is easily performed in time $O(n)$. The
space usage in internal memory is reduced to $O(\log n+\log \sigma)$ (plus
what is necessary to allow buffering for external memory).

Observe that in the reordering of values from rank to position order in the
previous section the part taking the most IM is step $2$. This is
$O(\sigma\log n)$ bits. This is again caused by keeping track of the $B$
function for each symbol of the alphabet while scanning the $\BWT$ to
compute an \textsc{LF} mapping. As described above we can perform this
\textsc{LF} mapping in EM while using $O(\log n+\log \sigma)$ in IM without
asymptotically using more space in EM or time. This leads us to the
following result.
\begin{theorem}
The succinct 2n bit PLCP representation for a string $s$ of length $n$ can,
given it's \texttt{BWT} and sampled suffix array of sampling rate $\mathcal{s}\in O(\log n)$, be constructed in worst cast time $O(n^2\log\sigma)$
and average time $O(n\log n\allowbreak\log\sigma)$ using $O(n\log \sigma)$ bits of space in EM and $O(\log\sigma +\log n)$ bits of space in IM.
\end{theorem}
\begin{figure}
\begin{algo}{PLCPexternal}{\BWT,n,\ISA}
\SET{(\texttt{Q},\texttt{S})}{(\emptyset,\textnormal{bit vector of $n$ \texttt{false} bits})}
\ACT{\texttt{Q}.\textsc{enque}((0,n))}
\DOWHILE{|\{i \mid \texttt{S}_i = \texttt{true}\}| < n}
\SET{\mathcal{Z}}{\textnormal{empty sequence}}
\DOWHILE{\texttt{Q}.\textsc{hasNext}()}
	\SET{(\texttt{$r_l$},\texttt{$r_h$})}{\texttt{Q}.\textsc{next}()}
	\COM{extract sub sequence of $\BWT$ for interval $[r_l,r_h)$}
	\SET{A}{\BWT_{r_l}\BWT_{r_l+1}\ldots \BWT_{r_h-1}}
	\ACT{\textnormal{sort }A\textnormal{ in time $O(|A|\log\sigma)$}}
	\COM{\textnormal{attach count to first occurence of each symbol and append to $\mathcal{Z}$}}
	\SET{\ell}{0}
	\DOWHILE{\ell < r_h-r_l}
		\SET{(h,a)}{(\ell+1,A[\ell])}
		\COM{find end of range for same symbol}
		\DOWHILE{h < r_h-r_l \textnormal{ and } A_h=a }
			\SET{h}{h+1}
		\OD
		\ACT{\mathcal{Z}.\textsc{append}((a,h-l))}
		\DOFORI{i}{1}{(h-\ell)-1}
			\ACT{\mathcal{Z}.\textsc{append}((a,0))}
		\OD
		\SET{\ell}{h}
	\OD
\OD
\ACT{\textnormal{sort $\mathcal{Z}$ by symbol component in time $O(n\log\sigma)$}}
\COM{construct bit vector $\mathcal{Z'}$ marking ranks which will get set in this round}
\DOFORI{r}{0}{n-1}
	\SET{(a,c)}{\mathcal{Z}_{r}}
	\SET{\mathcal{Z}'_{r}}{(c\ne 0 \textnormal{ and }S_r=\texttt{false})}
\OD
\COM{perform $\textsc{LF}^{-1}$ mapping on $\mathcal{Z'}$}
\ACT{\textnormal{inverse sort $\mathcal{Z}'$ using $\BWT$}}
\COM{activate ranks for this round}
\DOFORI{r}{0}{n-1}
	\COM{if rank $r$ not yet set but $\textsc{LF}(r)$ will be set in this round}
	\IF{\mathcal{Z'}_{r}=\texttt{true} \textnormal{ and }S_r=\texttt{false}}
		\ACT{\textnormal{activate $r$}}
	\FI
\OD
\ACT{\textnormal{increment count for active ranks}}
\COM{update $S$ and active set, construct intervals for next round}
\SET{\texttt{NQ}}{\emptyset}
\DOFORI{r_{l'}}{0}{n-1}
	\SET{(a,c)}{\mathcal{Z}_{r_l'}}	
	\IF{c\ne 0}
		\ACT{\textnormal{deactivate $r_l'$ and set $S_{r_l'}$}}
		\ACT{\texttt{NQ}.\textsc{enque}(r_l',r_l'+c)}
	\FI
\OD
\SET{\texttt{Q}}{\texttt{NQ}}
\OD
\end{algo}
\caption{Low internal memory variant PLCPexternal}
\label{lowimalg}
\end{figure}

%
%\vspace{-2mm}
\section{Improvement of Worst Case}
%\vspace{-2mm}
%
While on average our algorithm has a run time of $O(n\log n\log\sigma)$ as
the LCP values are $O(\log n)$ on average, we often see cases in
practice where, while most of the LCP values are small (in the order of $\log n$),
there are some significantly larger values as well. In this case an easy
adaption of our algorithm is to stop the computation of the $\texttt{PD}$
vector after a certain number of rounds (say $3\log n$) and compute the
missing values using the algorithm presented in \cite{DBLP:conf/wea/KarkkainenK14}.
This adaption can be performed using the following steps before reordering the $\texttt{PD}$ bit vector.
\begin{enumerate}
\item Erase all zero bits from the $\texttt{PD}$ bit vector corresponding to ranks which are still in the
active set. This removes incomplete values from $\texttt{PD}$ for
such ranks $r$ where $\texttt{LCP}[r]$ was not yet reached but
$\texttt{LCP}[\textsc{LF}(r)]$ was. This filtering takes time $O(n)$.
\item Compute a list $S_{im}$ (irreducible missing) of ranks $r$ in $S$ s.t.~$r=0$ or $r>0$ and $\texttt{BWT}[r-1]\ne \texttt{BWT}[r]$ in time $O(n)$ and space $O(n)$ bits of EM.
In the following let $n_{im}=|S_{im}|$.
\item Compute the list $S_{imlf}$ containing the ranks in $S_{im}$ and in addition for each rank $r\in S_{im}$ also $\textsc{LF}(r)$. 
This takes time $O(n\log\sigma)$ and space $O(n\log\sigma)$ in EM
where we use a scan over $\texttt{BWT}$ and a subsequent sorting by a symbol component as described above for computing the $\textsc{LF}$ function for a set of ranks.
This steps adds all ranks for the previous position of a rank in $S_{im}$, which we need for computing differences
between PLCP values for positions $p$ in $S_{im}$ and the respective previous positions $p-1$.
\item For each rank $r>0$ in $S_{imlf}$ add $r-1$ to $S_{imlf}$ in time $O(n)$.
We need these ranks for computing $\texttt{LCP}$ values because $\texttt{LCP}[r]$ is defined 
by comparing the suffixes at the ranks $r$ and $r-1$.
\item Convert $S_{imlf}$ to block code using $O(\log n)$ bits per rank in time $O(n)$ and space $O(n_{im}\log n)$ bits in EM.
\item Given a sampled inverse suffix array of sampling rate $\mathcal{s}\in O(\log n)$ use a method similar to
reordering the PLCP difference values above to annotate each rank in $S_{imlf}$ with the corresponding position
in time $O(n\log n\log\sigma)$ and space $O(n\log\sigma+n_{im}\log n)$ bits in EM.
\item Sort the resulting tuples by rank in time $O(n_{im}\log n)$ and space
$O(n_{im}\allowbreak \log n)$ bits in EM.
\item For each $(r,p)$ in the tuples s.t.~there is some tuple $(r-1,p')$ construct $(r,p=\texttt{SA}[r],r-1,p'=\texttt{SA}[r-1])$ in time
$O(n_{im})$ and space $O(n_{im}\log n)$ EM bits.
\item Annotate the tuples with the respective $\texttt{LCP}$ value between rank $r$
and $r-1$ stored in block code using a sparse version the algorithm presented in \cite{DBLP:conf/wea/KarkkainenK14}.
This requires the text $s$, which, if necessary, can be reconstructed from the $\texttt{BWT}$ and an inverse
sampled suffix array at sampling rate $\mathcal{s}\in O(\log n)$ in time $O(n\log n\log\sigma)$ and space $O(n\log\sigma)$ in EM.
Given $M\in O(n)$ words of IM (i.e.~$O(M\log n)$ bits) of IM this requires time
$O(\frac{n^2}{M\log_{\sigma}n} + n\log_{\frac{M}{B}}\frac{n}{B})$ using a
disk block size of $B$ words (see \cite{DBLP:conf/wea/KarkkainenK14}). Drop the $r-1$ and $p'=SA[r-1]$ components from the tuples.
\item Sort the tuples by position. Drop all tuples for positions $p$ s.t.~$p>0$ and there is no tuple for $p-1$.
For the rest replace the LCP component by the difference of the values for $p$ and $p-1$ if $p>0$.
\item Sort the tuples by rank (time $O(n_{im}\log n)$) and insert the computed values into the $\texttt{PD}$ bit vector (time $O(n)$).
\end{enumerate}
Using this hybrid algorithm we can obtain a trade off between the faster
worst case run time of the algorithm presented in \cite{DBLP:conf/wea/KarkkainenK14}
given sufficient IM and the reduced EM space usage of our algorithm presented
above. In this second stage of the hybrid algorithm we are generally only interested in computing
values for so called irreducible LCP values (cf.~\cite{DBLP:conf/cpm/KarkkainenMP09})
as only such values produce $0$ bits in the succinct PLCP vector.
The sum over all irreducible LCP values for any string of length $n$ is
bounded by $2n\log n$ (see \cite{DBLP:conf/cpm/KarkkainenMP09}). This bound
is reached for de Bruijn strings (cf. \cite{DBLP:conf/cpm/KarkkainenMP09}),
however in this setting each irreducible LCP value is $\Theta(\log n)$.
If we run the algorithm from the previous Section \ref{succinctlcpalgorithm}
for $O(\log^2 n)$ rounds, then all LCP values which remain unset must have a value of
$\Omega(\log^2 n)$, which means there are $O(\frac{n}{\log n})$ such values
and consequently the hybrid algorithm runs in worst case time $O(n\log^2 n\log \sigma)$ while using $O(n\log\sigma)$ space in EM and $O(\frac{n}{\log n})$ bits in IM.
\begin{theorem}
Given the \texttt{BWT} and sampled inverse suffix array of sampling rate $\mathcal{s}\in O(\log n)$ for a
string $s$ of length $n$ over an alphabet of size $\sigma$ the succinct permuted LCP array for $s$ can be
computed in time $O(n\log^2 n\log\sigma)$ while using $O(n\log\sigma)$ bits
of space in EM and $O(\frac{n}{\log n})$ bits of space in IM.
\end{theorem}
As the bound of $2n\log n$ for the sum over the irreducible LCP values of a
string is obtained for LCP values which are all of length $O(\log n)$ the
interesting question remains whether there is a smaller upper bound for the
sum of the irreducible LCP values when only LCP values in $\omega(\log n)$
are considered in the sum.
\section{Circular strings}
In this section we relax the original requirement of a unique terminator
symbol in $s$, i.e.~we no longer require that $s_{n-1} < s_{i}$ for all $i<n-1$.
Let $\hat{s}=\hat{s}_0\hat{s}_1\ldots$ be the infinite string defined by
$\hat{s}_i=s_{i\textnormal{ mod }n}$. Further let
$\hat{s}[i\dd]=\hat{s}_i\hat{s}_{i+1}\ldots$ for $i\geq 0$, i.e.~the suffix
of $\hat{s}$ starting from index $i$. We define that for two indices $i,j$
the relation $\hat{s}[i\dd] < \hat{s}[j\dd]$ holds if either there is some
$l$ s.t.~$\hat{s}_{i+l} < \hat{s}_{j+l}$ or $\hat{s}[i\dd]=\hat{s}[j\dd]$
and $i<j$. According to this definition we either have $\hat{s}[i\dd] <
\hat{s}[j\dd]$ or $\hat{s}[j\dd]<\hat{s}[i\dd]$ for $i\ne j$ and in
consequence there is a unique permutation 
$\hat{\texttt{SA}}=\hat{\texttt{SA}}_0,\hat{\texttt{SA}}_1,\ldots,\hat{\texttt{SA}}_{n-1}$ of
$0,1,\ldots,n-1$ s.t.~$\hat{s}[\hat{\texttt{SA}}_{i-1}\dd] < \hat{s}[\hat{\texttt{SA}}_{i}\dd]$ for $0< i < n$
and we can define $\hat{\texttt{BWT}}[i]=\hat{s}_{\hat{\texttt{SA}}_{i} + n - 1}$.
When defining a longest common prefix array for circular strings we face the
issue of identical suffixes even when they start at different indices and
thus infinite values in the array. These (infinite values) obviously occur in exactly such
cases when $s$ is an integer power of a string shorter than $s$ (i.e.~there is some string $w$ s.t. $s=ww\ldots
w$ which we write as $w^k$ if $s$ consists of $k$ copies of $w$ juxtaposed). This case
is easily detectable by scanning the BWT and determining whether there is some
$k$ dividing $n$ s.t.~for each $i$ in $0,1,\ldots,\frac{n}{k}-1$ we have
$\hat{\texttt{BWT}}[ik+0]=\hat{\texttt{BWT}}[ik+1]=\ldots=\hat{\texttt{BWT}}[ik+k-1]$. Figure
\ref{intpowalg} shows a linear time algorithm for detecting the maximum
period $p$ of $s$ s.t. $s=s[0\dd p-1]^{\frac{n}{p}}$.
% abcaabca
% ababb
\begin{figure}
\begin{algo}{detectPeriod}{\hat{\texttt{\texttt{BWT}}},n}
	\SET{(e,i)}{(\infty,0)}
	\DOWHILE{e > 1 \textnormal{ and }i < n}
		\SET{(j,c)}{(i+1,\hat{\texttt{\texttt{BWT}}}[i])}
		\DOWHILE{j < n \textnormal{ and }c=\hat{\texttt{\texttt{BWT}}}[j]}
			\SET{j}{j+1}
		\OD
		\IF{e=\infty}
			\SET{(e,i)}{(j-i,j)}
		\ELSE
			\SET{(e,i)}{(\textsc{gcd}(j-i,e),j)}
		\FI
	\OD
	\RETURN{\frac{n}{e}}
\end{algo}
\caption{Linear time algorithm for detecting maximum period $p$ s.t.~the
string of length $n$ underlying $\hat{\texttt{BWT}}$ equals $w^{\frac{n}{p}}$
for some word $w$
}
\label{intpowalg}
\end{figure}
For obtaining a meaningful LCP array for a string $s=w^e$ for $e>1$ we may
choose to shrink it's $\hat{\texttt{BWT}}$ array to that of a single base factor
$w$ by keeping every $e$'th symbol and discarding the symbols at
the other indices.

In the following we assume that $s$ is not an integer power of a word
shorter than $s$ and has length $n>1$, i.e.~$s$ contains at least two
different distinct symbols. As shown above this implies that for $0\leq i <
j < n$ there is always some $0\leq l < n$ s.t.~$\hat{s}_{i+l}\ne \hat{s}_{j+l}$.
In consequence there is a well defined array $\hat{\texttt{LCP}}=\hat{\texttt{LCP}}_0,\hat{\texttt{LCP}}_1,\ldots,\hat{\texttt{LCP}}_{n-1}$
given by $\hat{\texttt{LCP}}_0=0$ and $\hat{\texttt{LCP}}_i=l$ for $i=1,2,\ldots,n-1$ where
$l$ is the smallest number s.t~$\hat{s}_{\hat{SA}_{i-1}+l}\ne \hat{s}_{\hat{SA}_{i}+l}$.
Note that setting $\hat{\texttt{LCP}}_0=0$ is consistent with the scheme for the
other ranks as the suffixes at ranks $0$ and $n-1$ start with different
symbols, i.e.~the length of their longest common prefix is $0$. This also
guarantees that the $\hat{\texttt{LCP}}$ array contains the value $0$ at least once.
Based on the arrays $\hat{\texttt{SA}}$ and $\hat{\texttt{LCP}}$ we can
define the array $\hat{\texttt{ISA}}$ of length $n$ by $\hat{\texttt{ISA}}_{\hat{\texttt{SA}}_i} = i$
for $i=0,1,\ldots,n-1$ and
$\hat{\texttt{PLCP}}=\hat{\texttt{PLCP}}_0,\hat{\texttt{PLCP}}_1,\ldots,\hat{\texttt{PLCP}}_{n-1}$ 
by $\hat{\texttt{PLCP}}_{i} = \hat{\texttt{LCP}}_{\hat{\texttt{ISA}}_{i}}$.
The property of $\hat{\texttt{PLCP}}_{i} - \hat{\texttt{PLCP}}_{i-1} \geq -1$
still holds with the same arguments as in the non circular case, in fact
this can even be extended to $\hat{\texttt{PLCP}}_{0} - \hat{\texttt{PLCP}}_{n-1} \geq -1$
as the position $0$ has no special meaning in the circular case. Note
however that we loose one feature crucial for the $2n$ bit succinct PLCP
representation in the transition to circular strings and this is the
guarantee of $\hat{\texttt{PLCP}}_{n-1} = 0$ which stems from the unique
terminator symbol ensuring that no other suffix relevant for the computation
of $\hat{\texttt{LCP}}$ starts with the same symbol as the one at position $n-1$.
As an example consider the string \texttt{abbab} with the $\hat{\texttt{PLCP}}$ array $2,1,0,0,3$
which would translate to the bit vector $0001110100001$ of length $13 > 10=2n$.
Note that given $\hat{\texttt{SA}}$ and a select dictionary on the bit
vector we can correctly decode the respective $\hat{\texttt{LCP}}$ values,
however the vector is too long for the $2n$ bit bound. The reason for the
excessive length is precisely the fact that the $\hat{\texttt{PLCP}}$ array does
not end with a $0$ value. If we start off with the word \texttt{babba} which is a
rotation of \texttt{abbab} and consequently has the same
$\hat{\texttt{BWT}}$ then the $\hat{\texttt{PLCP}}$ array is rotated to
$3,2,1,0,0$ with the bit vector $0000111101$ of length $10=2n$. We chose
\texttt{babba} because it shifts the positions by $1$ from \texttt{abbab}
and thus moves the last $0$ at position $n-2$ in the $\hat{\texttt{PLCP}}$ array of
\texttt{abbab} to position $n-1$ in the array for \texttt{babba}. We can
obtain $\hat{\texttt{PLCP}}_i$ for \texttt{abbab} by decoding $\hat{\texttt{PLCP}}_{(i+1)\textnormal{ mod }n}$
for \texttt{babba} from the succinct PLCP bit vector for \texttt{babba}.
Suitable ranks $\hat{r}$ s.t.~$\hat{\texttt{LCP}}_{\hat{r}}=0$ can be found by checking the $\texttt{D}$ array.
Having chosen one such rank $\hat{r}$ we can deduce the respective position $\hat{p}$ by
using a sampled inverse suffix array and the BWT in time $O(n\log\sigma\log n)$ if
the sampling rate is $O(\log n)$ while using $O(n\log\sigma)$ bits in EM and
$O(\log\sigma+\log n)$ bits in IM.

For computing the succinct PLCP bit vector of a string using $\hat{\texttt{BWT}}$ and a sampled inverse suffix array observe that the
algorithm we presented in Section \ref{succinctlcpalgorithm} and \ref{redimsect} has no knowledge about
positions until it reaches the stage of reordering the values from rank to
position order. All the generated values are purely differential
(i.e.~$\hat{\texttt{PLCP}}_i - \hat{\texttt{PLCP}}_{(i+n-1)\textnormal{ mod }n}$
for $0\leq i<n$), in particular there is no special handling of position $0$.
The algorithm produces the bit vector $1110100001$ for the input \texttt{abbab}
which we need to rotate to $0000111101$ as described above to obtain correct
PLCP values while taking the employed position shift into account during decoding.
The hybrid algorithm can also be adapted for circular strings without
asymptotically modifying it's runtime or space usage. In step $9.$ we need
to take care of the fact that the comparison of two suffixes may extend
beyond the end of $s$. Due to our pre conditions however we can guarantee
that the longest common prefix of two different suffixes is always shorter
than $n$ symbols. This means that two runs over the set of blocks the text is
decomposed into in the original algorithm are sufficient, where in the
second run no more tuples are added but we only handle such tuples where the
comparison extends across block boundaries. When accessing the text we need
to use it's circular extension for comparisons. In step $10.$ we need to
handle the pair of positions $(n-1,0)$ if both positions are present.
Asymptotically we keep the same time bound for the hybrid algorithm as we extend the amount of work done in step 
$9.$ by a constant factor $2$ and in step $10.$ by a finite amount. This
gives us the following result.
\begin{theorem}
Given the circular $\hat{\texttt{BWT}}$ and sampled inverse suffix array of sampling rate $\mathcal{s}\in O(\log n)$ for a
circular string $\hat{s}$ deduced from a string $s$ of length $n$ over an alphabet of size $\sigma$ 
the succinct permuted LCP array for $\hat{s}$ can be
computed in time $O(n\log^2 n\log\sigma)$ while using $O(n\log\sigma)$ bits
of space in EM and $O(\frac{n}{\log n})$ bits of space in IM.
\end{theorem}
For the sake of this theorem the succinct permuted LCP array denotes the
shifted version plus respective position shift described above. If the
input string $s$ is an integer power of a shorter string $s'$ s.t.~$s'$ is
not itself an integer power of a shorter string, then the succinct permuted
LCP array is constructed using $s'$.

%\cite{mergebwtext}

%\bibliographystyle{abbrv}
\bibliography{article}

\end{document}